\documentclass[11pt]{article}
\usepackage{amssymb}
\usepackage{graphicx}
\usepackage{citesort}
\usepackage{setspace}

\begin{document}
\pagestyle{empty} \doublespacing

\begin{center}
  {\LARGE \bf Fast Dynamics of Glass-forming Liquids investigated by Dielectric Spectroscopy: a Comparison}\\
 \vspace{5mm} U. Schneider, P. Lunkenheimer, R. Brand, and
A. Loidl\\ {\it Experimentalphysik V, Institut f\"{u}r Physik, Universit\"{a}t Augsburg,\\ D-86135 Augsburg, Germany}
\end{center}
\vspace{5mm}

\section*{Motivation}

Understanding the mechanism of the glass transition is one of the most challenging tasks of modern condensed matter
physics. Despite the ample experimental and theoretical work on this subject there is still no universally accepted
view thoroughly describing all physical aspects. Among the experimental approaches used to investigate this phenomenon,
dielectric spectroscopy is a well established method probing the rotational dynamics of dipolar molecules and the
translational dynamics of charged particles.

Our group has access to an extraordinary wide range of frequencies from 10$^{-6}$ to 10$^{14}$\,Hz almost covering 20
decades continuously. The various experimental setups include a home-made time-domain spectrometer (10\,$\mu$Hz $\leq
\nu \leq 1$\,kHz), commercially available autobalance bridges ($20$ Hz $\leq \nu \leq 20$ MHz), and radio-frequency and
microwave setups using coaxial reflection and transmission techniques ($1$ MHz $\leq \nu \leq 30$ GHz). At $40$ GHz
$\leq \nu \leq 1.2$ THz a quasi-optical submillimeter-wave spectrometer is employed measuring the complex transmission
coefficient. Higher frequencies (450\,GHz $\leq \nu \leq 10$\,THz) are investigated with a commercially available
Fourier-transform infrared spec\-tro\-meter.

We applied these techniques to investigate the glass transition in the molecular glass formers glycerol ($T_g =185$\,K,
hydrogen-bonded network) \cite{Lunkigly,Lunkiorl,Lunkikyo,Lunkibos,Sch98} and propylene carbonate (PC) ($T_g=160$\,K,
van-der-Waals glass) \cite{Lunkigly,Lunkiorl,Lunkikyo,Lunkibos,Sch99} and in the ionic melts
[Ca(NO$_3$)$_2$]$_{0.4}$[KNO$_3$]$_{0.6}$ (CKN) \cite{Lunkigly,Lunkiorl,Lunkikyo,Lunkibos,Lun97} and
[Ca(NO$_3$)$_2$]$_{0.4}$--[RbNO$_3$]$_{0.6}$ (CRN) (for both: $T_g =333$\,K). The frequency dependent complex
dielectric permittivity $\varepsilon{}^*=\varepsilon{}'-i\varepsilon{}''$ obtained from the experiments allows for an
investigation of a variety of dynamic processes known to be present in glass-forming materials including
$\alpha$-process, fast $\beta$-processes, and microscopic response (boson peak). We are able to trace the dynamics of
the $\alpha$-relaxation from almost total arrest near the glass temperature $T_g$ up to temperatures in the liquid
state where it starts to merge with the microscopic response.

Of special interest is the high-frequency r\'egime in the GHz -- THz range, which allows for an investigation of the
possible fast processes, that have been predicted by various theoretical approaches to be inherent to supercooled
liquids. We provide a detailed analysis of these processes using the predictions of the mode coupling theory (MCT) of
the glass transition. \cite{Goe92} This theory developed in the last 20 years describes the glass transition as a
dynamic phase transition at $T_c > T_g$. It is still controversially discussed and until recently could not be tested
properly with dielectric spectra because of the limited frequency range available in typical dielectric experiments.
The measurements of $\varepsilon{}''(\nu)$ in the far-infrared region at $\nu\approx1$ THz give access to the r\'egime
of the boson peak, known already from neutron and light scattering experiments with which our data will be compared.

\section*{Results and Analysis}

\begin{figure}[tbp]
  \centering
  \begin{minipage}[t]{0.48\textwidth}
  \centering
  \includegraphics[clip,width=7.7cm]{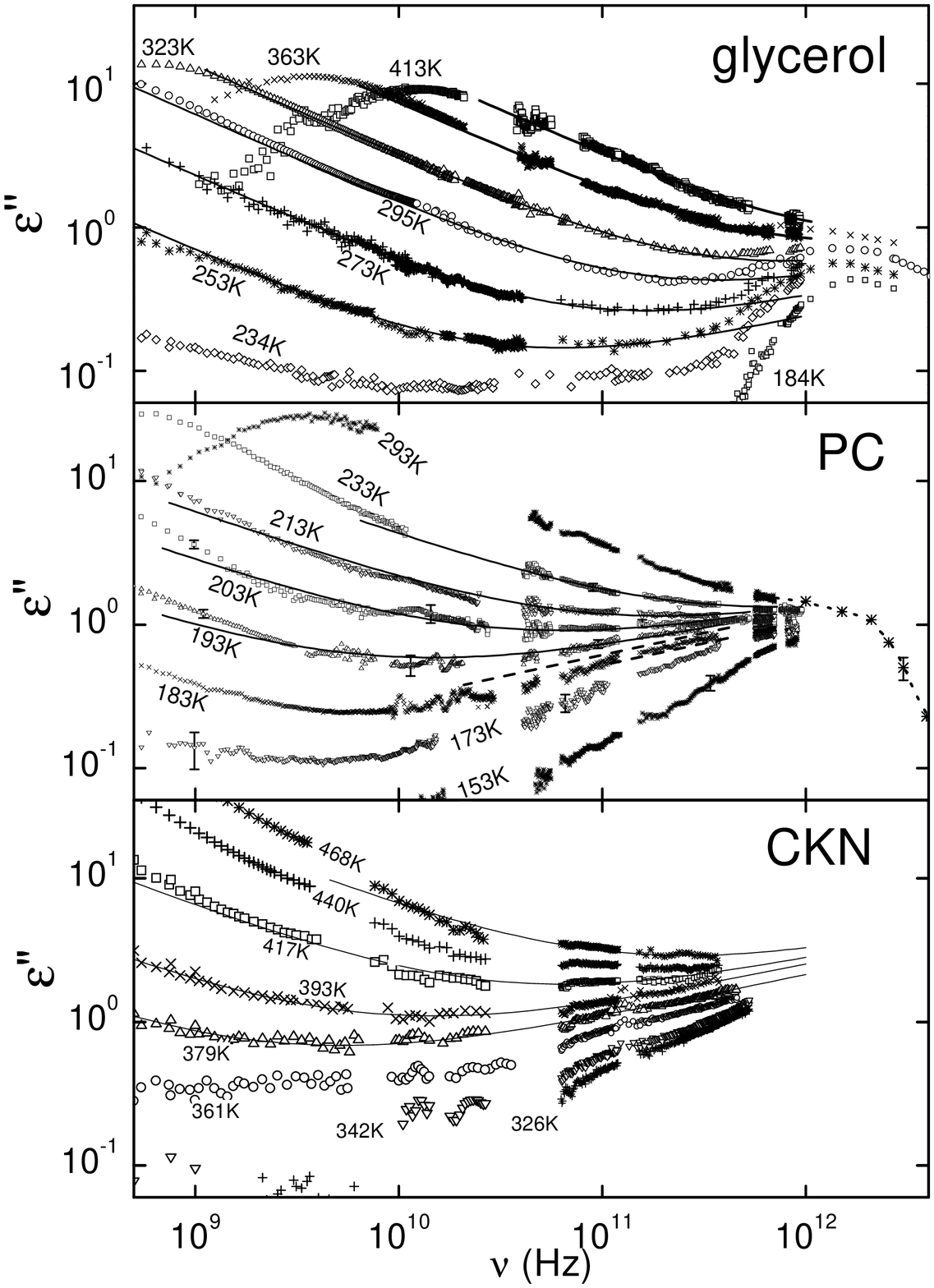}
  \sf \caption{High frequency $\varepsilon{}''(\nu)$ spectra of glycerol, PC and CKN for various temperatures. The lines are fits according to the predictions of the idealized MCT \protect\ref{eq:mct} (for parameters see text).}
  \rm \label{abb:fig1}
  \end{minipage}\hfill
  \begin{minipage}[t]{0.48\textwidth}
  \centering
  \includegraphics[clip,width=7.8cm]{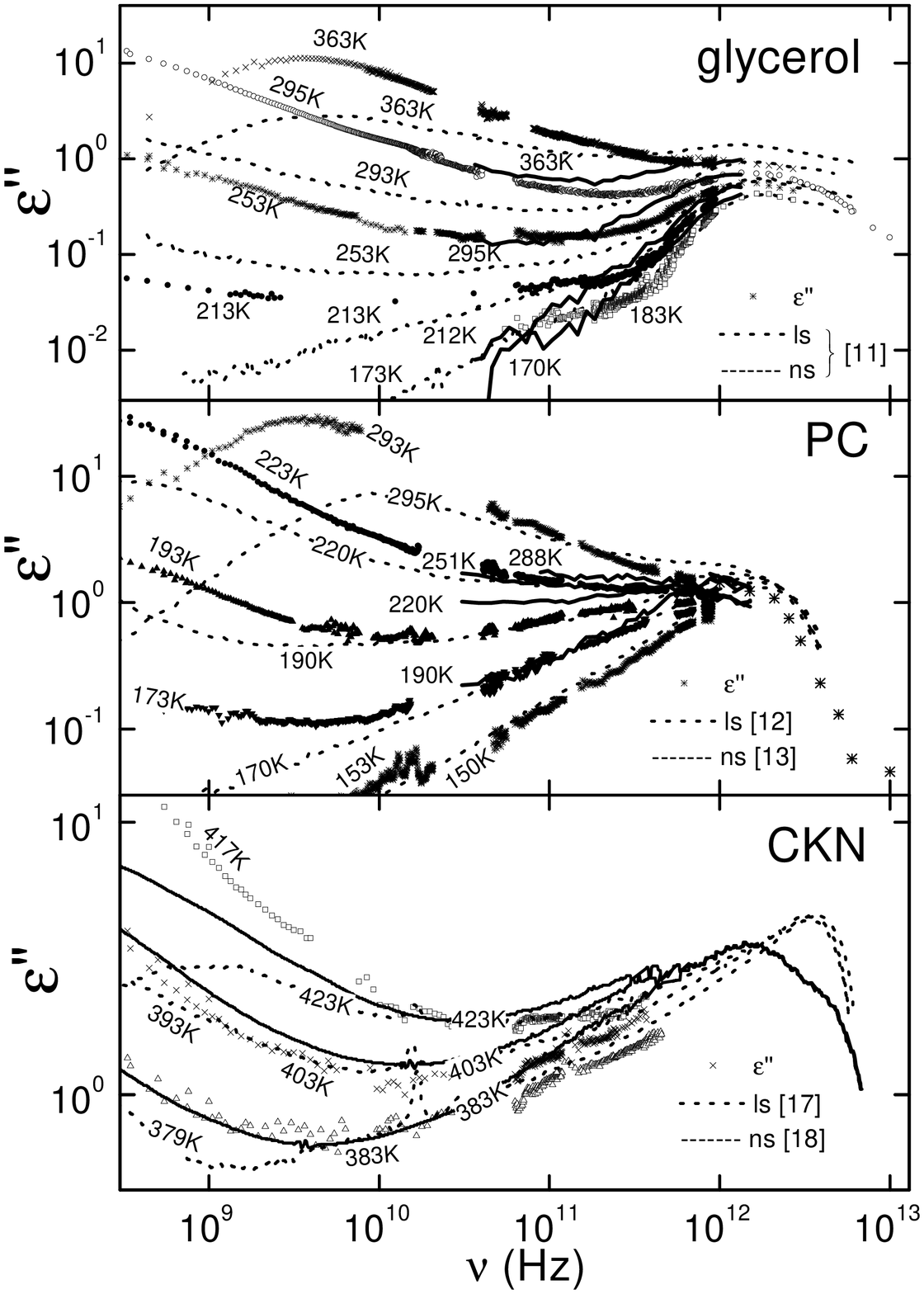}
  \sf \caption{High frequency $\varepsilon{}''(\nu)$ spectra of glycerol, PC and CKN for various temperatures. The spectra with symbols are the dielectric data of $\varepsilon{}''(\nu)$, the lines are $\chi ''(\nu)$ as calculated from light and neutron scattering data.}
  \rm \label{abb:fig2}
  \end{minipage}
\end{figure}

Figure~\ref{abb:fig1} shows the high frequency part of the dielectric loss spectra for glycerol
\cite{Lunkigly,Lunkiorl,Lunkikyo,Lunkibos,Sch98}, PC \cite{Lunkigly,Lunkiorl,Lunkikyo,Lunkibos,Sch99} and CKN
\cite{Lunkigly,Lunkiorl,Lunkikyo,Lunkibos,Lun97}. We observe a temperature dependent $\alpha$-peak, a shallow minimum
region and the boson-peak (not in CKN). According to the so-called idealized MCT \cite{Goe92}, above $T_{c}$, the
minimum region can be approximated by the interpolation formula:
 \begin{equation} \varepsilon{}''(\nu)=\frac{\varepsilon{}''_{\min }}{(a+b)}\left[a\left( \frac{\nu}{\nu_{\min }}\right)^{-b}+b\left(\frac{\nu}{\nu_{\min }}\right)^{a}\right]\label{eq:mct}\end{equation}
$\nu _{\min}$ and $\varepsilon{}''_{\min}$ denote the position the minimum. The exponents $a$ and $b$ are temperature
independent and are constrained by the exponent parameter $\lambda =\Gamma ^{2}(1-a)/\Gamma (1-2a)=\Gamma
^{2}(1+b)/\Gamma (1+2b)$ where $ \Gamma $ denotes the Gamma function. This formula restricts the exponent $a$ to values
below $0.4$, i.e.~a significantly sublinear increase of $\varepsilon{}''(\nu )$ at frequencies above $\nu_{\rm min}$ is
predicted. The critical temperature $T_c$ should manifest itself in the temperature dependence of the
$\varepsilon{}''(\nu )$-minimum. For $T>T_c$ MCT predicts the following relations: $\nu _{\rm min}\sim
(T-T_c)^{1/(2a)}$ and $\varepsilon _{\rm min}''\sim (T-T_c)^{\frac{1}{2}}$. Additionally, MCT predicts critical
behaviour also for the $\alpha$-relaxation: the peak frequency follows $\nu_{max} \sim (T-T_c)^\gamma$ with
$\gamma=1/(2a)+1/(2b)$. In the case of the ionic conductors we chose the imaginary part of the dielectric modulus
$M''(\nu)={\rm Im}\{1/\varepsilon{}^*(\nu)\}$ to determine $\nu_{max}$ for the $\alpha$-relaxation since the maxima in
$\varepsilon{}''(\nu)$ are hidden by large conductivity contributions. This procedure is commonly used for the
evaluation of dielectric data on ionic conductors \cite{Moyni} and justified by the finding that the results for
$M''(\nu)$ follow closely those obtained by mechanical experiments.\cite{Pimenov}

The solid lines in figure~\ref{abb:fig1} are fits of the minimum region of $\varepsilon{}''(\nu)$ with
equation~\ref{eq:mct}. For glycerol reasonable fits are obtained for $\nu$\,\,${\scriptstyle\lesssim}$\,\,$\nu_{\rm
min}$ with $\lambda =0.705$, $a=0.325$, $b=0.63$. \cite{Lunkigly,Lunkibos} These values agree reasonably with the
results from other techniques.\cite{Wut94} At high frequencies the fits are limited by an additional steeper increase.
It may be argued that these deviations are due to vibrational contributions (the so-called boson-peak) which are not
included in the idealized version of MCT. A different behaviour is observed in the PC spectra: vibrational
contributions seem to be of less importance and we find a consistent description of the $\varepsilon{}''(\nu)$-minima
at $T\geq 193$ K using $\lambda =0.76$ which implies $a=0.29$ and $b=0.5$. The obtained value of the exponent parameter
$\lambda $ is consistent with the results from various other measurement techniques \cite{Du,Ohl,Berg}. The strong
boson-peak in glycerol is in accord with the findings of Sokolov {\it et al.}\,\cite{Sok} that the amplitude ratio of
boson peak and fast process is largest for 'strong' glass formers \cite{Angell}, glycerol being much stronger than PC.
The high frequency CKN spectra above 379\,K can be fitted very well using equation~\ref{eq:mct}, yielding $\lambda
=0.76$, $a=0.3$, $b=0.54$.~\cite{Lun97} The parameters are in good agreement to those obtained from light scattering
experiments.~\cite{Li}. For CRN (spectra not shown) we obtained the following set of parameters: $\lambda =0.91$,
$a=0.2$, $b=0.35$.~\cite{Lun97}

\begin{figure}[tbp]
  \centering
  \begin{minipage}[t]{0.7\textwidth}
  \centering
  \includegraphics[angle=-90,clip,width=12cm]{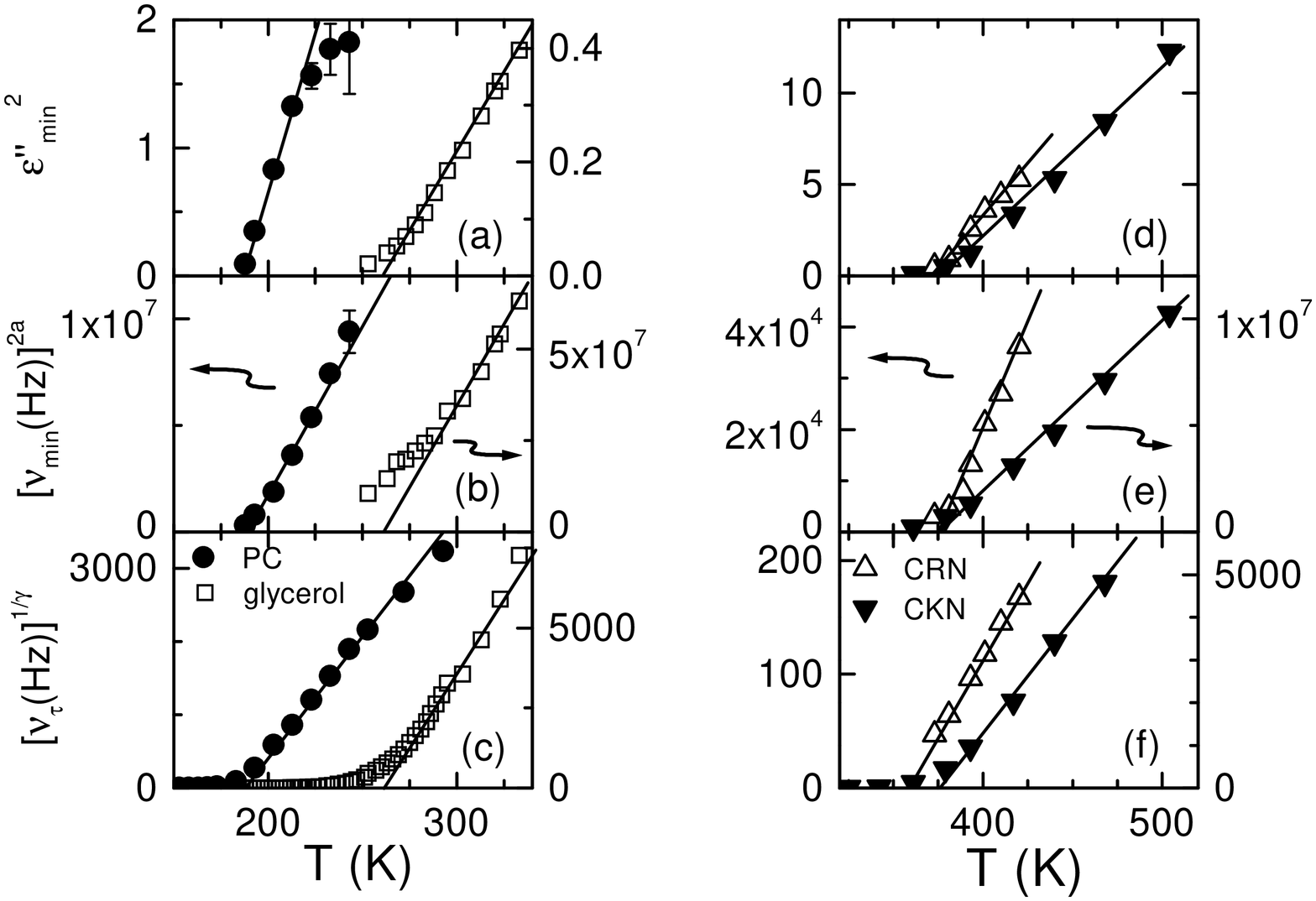}
  \end{minipage}\quad
  \begin{minipage}[t]{0.25\textwidth}
  \centering
  \sf \caption{Plots of the critical dependences for the position of the minimum for glycerol and PC[(a): height
$\varepsilon{}''_{\min}$, (b): frequency $\nu_{\min}$] and (c) the $\alpha$\,-peak; (d), (e) and (f) are the same plots
for CKN and CRN. The lines are drawn according to MCT as linear extrapolations to the corresponding $T_c$.}
\label{abb:fig3} \rm
  \end{minipage}
\end{figure}

The critical temperature $T_c$ should manifest itself in the temperature dependence of the
$\varepsilon{}''(\nu)$-minimum and the $\alpha$-peak (see above). Figure~\ref{abb:fig3} presents minimum amplitude and
frequency and the $\alpha$-peak frequency in representations that lead to straight lines, extrapolating to $T_c$, if
the predicted critical behaviour is obeyed. For all materials the sets of parameters can be described consistently with
$T_c\approx 262$\,K for glycerol, 187\,K for PC, 375\,K for CKN and 365\,K for CRN. These values lie in the same region
as the $T_c$'s obtained from other techniques.~\cite{Wut94,Du,Ohl,Berg,Li,Knaak} The deviations of the data from the
predicted critical behaviour seen near $T_c$ can be ascribed to hopping processes considered in extended versions of
MCT.~\cite{Goe92} Especially for glycerol some uncertainties for $T_c$ follow as a consequence of the choice from which
temperature range the extrapolation is made.

Finally we want to compare our spectra to those obtained from light and neutron scattering experiments. In
figure~\ref{abb:fig2} the dielectric spectra are plotted as symbols, the dotted lines are the imaginary part of the
susceptibility $\chi_{ls}''(\nu)$ calculated from light scattering and the solid lines are $\chi_{ns}''(\nu)$ from the
neutron scattering data taken from the literature~\cite{Wut94,Du,Ohl,Li,Knaak}. The scattering spectra, given in
arbitrary units, are scaled to match the $\varepsilon{}''(\nu)$ data in the boson-peak r\'egime. Comparing the three
materials in figure~\ref{abb:fig3} we observe the following universal characteristics: (1) The ratio of the structural
processes ($\alpha$-relaxation) and the boson-peak is largest in the dielectric and smallest in the neutron scattering
data. This is also found in molecular dynamics simulations of ortho-terphenyl~\cite{Wahn} and of a system of rigid
diatomic molecules.~\cite{Schsim} (2) The frequency of the $\alpha$-peak (where observable) is higher in the light
scattering data compared to the dielectric data. (3) The minimum in $\varepsilon{}''(\nu)$ only coincides with the
minima of the scattering data in CKN; for both glycerol and PC there are differences in the position of $\nu_{\min}$.

The differences in the spectra are caused by the different probes, each method couples to. A possible explanation of
the underlying microscopic processes was given considering the different dependencies of the probes on orientational
fluctuations.~\cite{Leb} Additionally, the MCT was recently generalized to molecular liquids with orientational degrees
of freedom~\cite{Schkug} thereby providing an explanation for the different ratios of $\alpha$- and boson-peak
amplitude for the different probes. In addition, very recently MCT was successfully applied to simultaneously describe
both the present dielectric and light scattering data~\cite{Li} of PC by means of a schematic model.~\cite{voigtmann}

\section*{Acknowledgements}

Our gratitude is directed to A. Maiazza for preparing the ionic conducting materials and A. Pimenov, Yu. Goncharov, B.
Gorshunov and M. Dressel for help with installing the submillimeter-wave spectrometer and performing some of the
measurements. This work was supported by the Deutsche Forschungsgemeinschaft, Grant-No. LO264/8-1 and the BMBF,
contract-No. 13N6917.

\end{document}